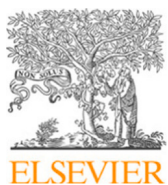

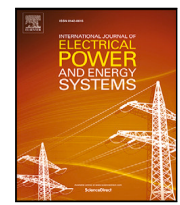

# Power Ramp-Rate Control via power regulation for storageless grid-connected photovoltaic systems

Jose Miguel Riquelme-Dominguez [a,*], Francisco De Paula García-López [b], Sergio Martinez [a]

[a] Electrical Engineering Department, Escuela Técnica Superior de Ingenieros Industriales, Universidad Politécnica de Madrid, C/ José Gutiérrez Abascal, 2, Madrid, 28006, Spain
[b] Department of Electrical Engineering, Escuela Técnica Superior de Ingeniería, University of Seville, C/ Américo Vespucio, Sevilla, 41092, Spain

## ARTICLE INFO

## ABSTRACT

Photovoltaic Power Ramp-Rate Control (PRRC) constitutes a key ancillary service for future power systems. Although its implementation through the installation of storage systems or irradiance sensors has been widely investigated, fewer studies have explored the power curtailment approach. The latter lacks efficiency, as it voluntarily produces power discharges, yet it is a cost-effective solution in terms of capital expenditures. This paper proposes a novel storageless and sensorless photovoltaic PRRC for grid-connected applications in which the photovoltaic power, rather than the voltage, is the controlled magnitude. The aforementioned contribution makes the effective tracking of the power ramp-rate limit possible compared to the existing methods in the literature. The method is assisted by a real-time curve-fitting algorithm that estimates the Maximum Power Point while operating suboptimally. Thus, no direct temperature or irradiance measurement systems are needed. The validation of the proposed PRRC strategy has been tested by simulation and compared to another approach available in the literature, considering real-field highly variable irradiance data. Experimental validation of the proposed strategy has been performed in real time via Controller Hardware-in-the-Loop.

## 1. Introduction

The displacement of conventional generation by renewable sources raises several issues related to power system stability. In fact, as a consequence of high renewable penetration, power systems present reduced inertia, which increases the risk of suffering from voltage rise, frequency deviations and output power fluctuations [1,2]. The latter issue is mainly due to variations in the primary energy source: solar irradiance in solar photovoltaic (PV) systems. These variations can be as fast as the speed of passing clouds [3].

It is generally accepted that a large geographical distribution of PV systems may reduce these fluctuations [1,4], however, in [5], it was observed that a large photovoltaic system of 1.6 megawatts can exhibit output power fluctuations exceeding 50% of its rated capacity in less than 10 s. For this reason, different Transmission System Operators (TSOs), from small to large power systems, are starting to demand Power Ramp-Rate Control (PRRC) capability from photovoltaic generators [6]. Table 1 summarizes power ramp-rate specifications in some power systems with different topology, robustness and renewable penetration level. It is worth mentioning the case of Hawaii, where not only 1-minute ramp-rate limit is considered. Specifically, Hawaiian Electric Company (HECO) sets an additional ramp-rate limit of 1 MW

### Table 1
Photovoltaic PRRC requirements by different TSOs.

| TSO | Country | Ramp-up | Ramp-down |
|---|---|---|---|
| PREPA | Puerto Rico | 10%/min | 10%/min |
| EirGrid | Ireland | 30 MW/min | – |
| HECO | Hawaii | 2 MW/min | 2 MW/min |
| German TSOs | Germany | 10%/min | 10%/min |

scanned every 2 s. Table 1 highlights the need to develop flexible algorithms that control the photovoltaic power ramp-rate in order to meet the requirements imposed by TSOs.

A large number of existing studies in the broader literature have examined the incorporation of Battery Energy Storage Systems (BESS) [7–9] to achieve PRRC. The main advantage of this solution is the optimization of the power generation, as no energy is wasted apart from conversion losses. However, the use of BESS requires advanced communication systems, the lifetime of batteries is still reduced and the overall cost of the PV system is considerably increased [10].

A different approach consists in the incorporation of sky-camera-based [11] or ground-based sensors [12] for predictive control of passing clouds. While the former has to deal with the shadow map at the ground level, thus affecting the effectiveness of the method, the latter can achieve high average accuracy. However, both require additional equipment and more complex communications, which may increase the capital investment of the system.

Recent studies [13–15] have investigated an alternative approach for photovoltaic PRRC consisting of power curtailment methods. Although the power curtailment approach implies energy waste, it requires lower capital investments which makes it an alternative to be considered [16,17], taking into account the particularities of each installation. Furthermore, power curtailment and storage are not necessarily mutually exclusive solutions, as in some situations the combination of both produces the best option from the economic point of view [18].

Reference [13] modified the Maximum Power Point Tracking (MPPT) algorithm by shifting the perturbation direction of the Perturb and Observe (P&O) algorithm when the PV power ramp-rate is bigger than a threshold. Nevertheless, in highly variable irradiance conditions, it may help the PV system to avoid MPPT drift, resulting in an ineffective strategy for these conditions.

Ref. [14] improved the previous method with a two-step strategy. Firstly, the power ramp-rate is determined by a Ramp-Rate Measurement (RRM) calculation and then the power ramp-rate is limited in subsequent steps by perturbing the operation voltage to the left of the Maximum Power Point (MPP) in the power–voltage (P–V) curve. This technique is simple and straightforward, but some considerations may be taken into account. As the MPPT algorithm is the P&O one, the RRM is not only affected by the filtering value $n$, but also by the perturbation size $V_{step}$. In addition, the controlled variable is the PV voltage rather than the PV power, which hinders the tracking of the maximum power ramp-rate allowed.

In [15], an alternative approach was recently presented. In this technique, the PV system is operated at a sub-optimal voltage, whose value is provided by the system operator. Then, with a certain frequency, the power ramp-rate is measured. If the measured ramp-rate exceeds the limit, the operating voltage is pushed to the left of the MPP, and finally, the operation voltage is regulated step by step to satisfy the power ramp-rate constraint. This method provides the remarkable possibility of partially limiting the ramp-down events. However, it is not solved how the system operator determines the value of $V_{opt}$. Furthermore, PRRC based on PV voltage control is less accurate due to the lack of knowledge of the real-time P–V characteristic.

In this paper, a novel power ramp-rate control is presented based on PV power regulation, rather than PV voltage. More precisely, according to the definitions of the International Electrotechnical Vocabulary [19], the controlled variable is the PV power instead of the PV voltage. To the knowledge of the authors, no prior studies have examined this procedure in PRRC applications. The proposed algorithm has two switched modes of operation: PRRC and MPPT. In steady-state irradiance conditions, the PV system is operated at a curtailed power level ($\Delta P$), following the power reference commanded by the system operator. A curve-fitting algorithm provides an estimation of the MPP and the actual power reserves in real-time without the need of using direct temperature or irradiance measurements. In rapidly changing irradiance conditions, PV power is regulated according to the maximum power ramp-rate permitted until the active power reserve is over and the MPPT mode is enabled.

The control of the PV power rather than the voltage provides two benefits compared to existing methods in the literature: the PV system reacts inherently to changes in irradiance in order to maintain the reference power and the maximum power ramp-rate can be effectively tracked, as simulation results and experimental validation exhibit.

In summary, the main contributions of this work are:

1. A novel storageless photovoltaic PRRC strategy is introduced, in which the PV system maintains active power reserves in order to smooth irradiance fluctuations.

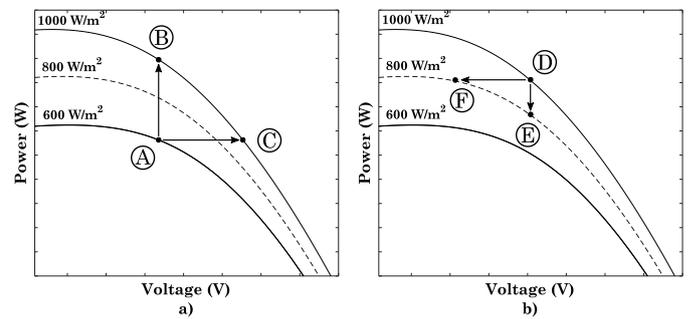

**Fig. 1.** Voltage vs. power control in case of: (a) irradiance increase and (b) irradiance decrease.

2. In contrast with previous methods for PV-PRRC, the presented strategy directly controls PV power instead of PV voltage, which is specifically interesting with high-demanding irradiance profiles, as results exhibit.

3. Experimental validation of the proposed technique has been performed through Controller Hardware-in-the-loop, therefore proving that the strategy is suitable for real-time implementation.

Finally, it is important to mention that the contributions of this paper are conceived for the case of low-cost systems, where there is no energy storage or additional sensor equipment. What is proposed here can be a complement to the existing technology in future power systems.

This paper is organized as follows: Section 2 presents the PV power control. Section 3 explains the proposed PRRC algorithm. Section 4 shows the simulation results. In Section 5, the experimental validation is presented. Finally, Section 6 draws the main conclusions of the work.

## 2. Photovoltaic power and voltage control

### 2.1. Voltage vs power control

Traditionally, in grid-connected photovoltaic systems, PV voltage has been used as the control objective for different control purposes, such as the implementation of MPPT algorithms [20]. However, in PRRC applications, PV voltage control may introduce undesired power peaks inconsistent with the power ramp-rate limitation. Fig. 1 illustrates the above statement.

In Fig. 1(a), when the operating point is point A, if PV voltage is the controlled variable, a sudden irradiance increase from 600 W/m² to 1000 W/m² moves the operating point towards point B resulting in an abrupt change in PV power. Conversely, if PV power is regulated, the control system reacts inherently when irradiance increases, so the power delivered by the PV system is constant (A → C). Similar reasoning can be used when irradiance drops. In Fig. 1(b), let us suppose the operating point is point D and irradiance falls from 1000 W/m² to 800 W/m². A voltage-regulated PV system reacts maintaining the operating voltage, that is, moving from point D to point E. On the other hand, if PV power is controlled, the working point is moved from D to F. In this way, the control system can avoid abrupt changes in the photovoltaic power. It should be noted that, for major drops of irradiance, it is possible that the PV system cannot maintain the PV power (the power reference is greater than the MPP). For example, if irradiance is reduced from 1000 W/m² to 600 W/m² in Fig. 1(b). The PRRC algorithm must activate the MPPT mode in this situation as there is no available power reserve.

From the above analysis, it is clear that PV power regulation is more suitable for PRRC strategies.

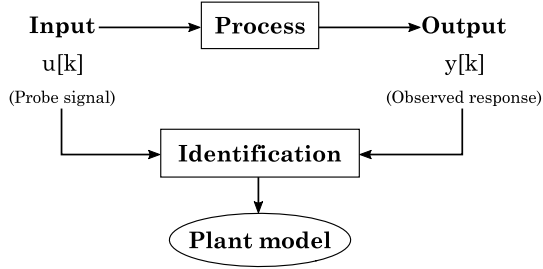

**Fig. 2.** Plant model identification diagram.

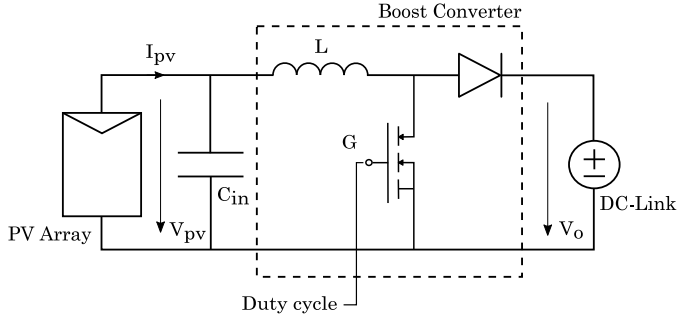

**Fig. 3.** Photovoltaic system under study.

**Table 2**
PV system parameters.

| Parameter | Value |
| --- | --- |
| $P_{pv0}$ | 2040 W |
| $C_{in}$ | 470 μF |
| $L$ | 500 μH |
| $V_0$ | 700 V |
| $f_s$ | 20 kHz |

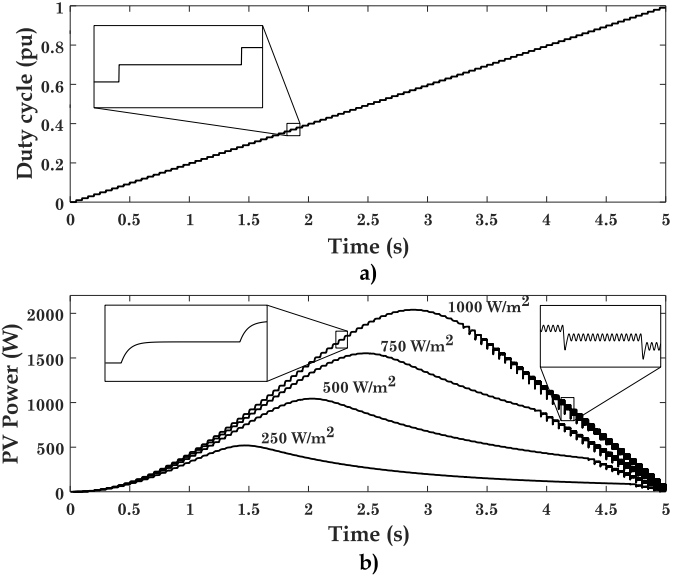

**Fig. 4.** PV power response (b) to duty cycle changes (a).

## 2.2. PV system under study

Before describing the PRRC proposed in this work, let us introduce the particular PV system that has been used to illustrate the technique and to test its implementation.

### 2.2.1. System identification

It is worth noting that photovoltaic power control is not a novel idea, as it has been implemented in [21] to get the PV system to maintain active power reserves. The novelty of the present study lies in the use of PV power control for PRRC applications. In [21], the PV power controller is designed according to the system model linearization. However, the linearization of the system requires the computation of derivatives and sophisticated variables such as the dynamic resistance. In the present work, the plant model has been obtained by direct identification of a continuous-time model from sampled data [22,23]. Fig. 2 summarizes the identification process.

This procedure treats the plant as a black-box, where a known input signal $u[k]$ is applied to the system and its response $y[k]$ is collected. The objective is to find a transfer function for the plant model that, applying the same input $u[k]$, generates an output $\hat{y}[k]$ that satisfies Eq. (1).

$$min \ \varepsilon = \sqrt{\frac{1}{N}\sum_{k=1}^{N}\left(y[k]-\hat{y}[k]\right)^2} \qquad (1)$$

where $\varepsilon$ is the Root Mean Square Error (RMSE) between the real plant and the estimated model.

System model identification is a general technique suitable for a large number of industrial processes. In this particular case, it is used for the identification of the PV system plant depicted in Fig. 3. As can be seen, it is composed of a PV array, a DC–DC boost converter and a voltage source acting as the DC-Link between conversion stages. This simplified model is widely used in the literature for grid-connected photovoltaic systems as the photovoltaic control is commonly made at the first conversion stage [15,20,21,24]. Table 2 lists the main parameters of the system used to validate the proposed technique.

For the system identification process, the input signal to the PV system is the boost converter duty cycle, whose value is a real positive number within the interval [0,1], and the output is the PV power. As the dynamic response of the system varies with the operating point, the step response needs to be evaluated for all possible input levels. To do so, a series of tests are carried out. They consist in the determination of multiple step responses, concatenated one after the other as Fig. 4 shows. Fig. 4(a) represents the time series of the step inputs to the system and Fig. 4(b) depicts several time series of responses to these steps with different irradiance levels.

The PV power output depends mainly on the incident irradiance and the ambient temperature. As the temperature dynamics is much slower than the irradiance one, the temperature is kept constant for the system identification process. This is a reasonable assumption on the time scale of interest in PRRC applications. Regarding irradiance, several levels must be considered for a correct identification. For this particular system under study, four irradiance levels have been taken into account: 1000, 750, 500 and 250 W/m². The duty cycle is varied from 0 to 1 in steps of 0.01 pu in periods of 0.05 s and the output PV power is recorded. The zoomed parts detailed in Fig. 4 manifest that there are clearly two different zones of operation: for low values of the duty cycle, i.e., from open-circuit voltage up to beyond MPP, the system behaves as a first-order system, and for high values of the duty cycle, i.e., near the short-circuit current, the system response is oscillatory. This reveals that the operation at the right-hand side of the P–V curve is more appropriate for PV power control.

In order to obtain the model of the PV system, transfer function identification is carried out with part of the data of Fig. 4, considering only those that belong to the right-hand side of the power-voltage curve, i.e., from the open-circuit voltage to the MPP. The identification has been performed with the embedded MATLAB function *tfest*, available in the System Identification Toolbox [25]. The transfer function to be estimated is in the form of Eq. (2), which corresponds to a first-order system.

$$G(s) = \frac{K}{\tau s + 1} \qquad (2)$$

**Table 3**
Transfer function identification summary.

| Irradiance (W/m²) | N° of TF | Estimation fit (%) |
|---|---|---|
| 1000 | 57 | 99.22 |
| 750 | 49 | 99.14 |
| 500 | 40 | 99.01 |
| 250 | 29 | 98.80 |

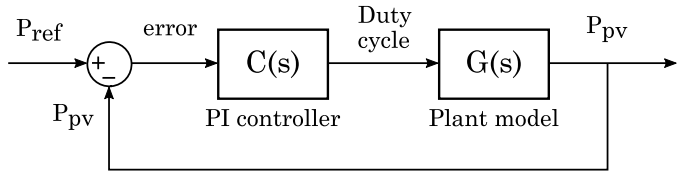

**Fig. 5.** Closed-loop control of PV power.

**Table 4**
A sample of six identified open-loop transfer functions with diverse pole locations.

| Sample | Open-loop transfer function |
|---|---|
| #1 | $C(s)G_1(s) = \frac{163.3s+8165}{s^2+207.2s}$ |
| #2 | $C(s)G_2(s) = \frac{343.2s+17160}{s^2+355.3s}$ |
| #3 | $C(s)G_3(s) = \frac{427.8s+21388}{s^2+259.5s}$ |
| #4 | $C(s)G_4(s) = \frac{99.82s+4991}{s^2+159.7s}$ |
| #5 | $C(s)G_5(s) = \frac{133.2s+6661}{s^2+323.1s}$ |
| #6 | $C(s)G_6(s) = \frac{519.7s+25984}{s^2+272.6s}$ |

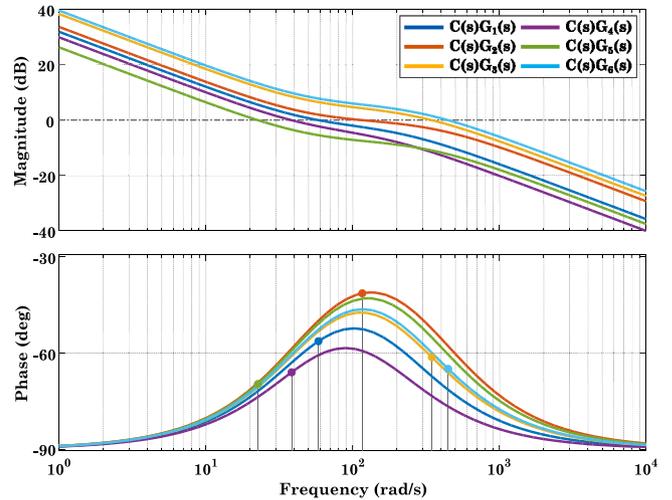

**Fig. 6.** Bode plots of the selected six open-loop transfer functions $C(s)G(s)$.

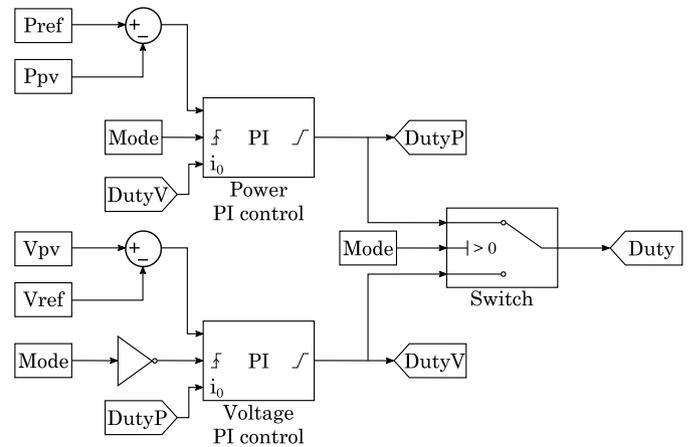

**Fig. 7.** Power-voltage control diagram.

**Table 5**
PI controllers parameters.

| Parameter | $P_P$ | $I_P$ | $P_V$ | $I_V$ |
|---|---|---|---|---|
| Value | $4 \times 10^{-4}$ | $2 \times 10^{-1}$ | $1 \times 10^{-2}$ | $5 \times 10^{0}$ |

Table 3 summarizes the results of the identification process. As can be seen, the estimation fit is well above 98% for the levels of irradiance considered. This study confirms that the system behaves as a first-order system in the operation range examined. It should be pointed out that the number of estimated transfer functions varies depending on the irradiance. This is due to the fact that, when irradiance is higher, the PV system reaches the MPP at a greater value of the duty cycle.

*2.2.2. System controller*

Once a set of transfer functions is identified, the next step consists of designing a unique PI control, in order to get the closed-loop control scheme of Fig. 5 to be stable in all possible conditions. The control system first compares the PV power with the power reference, which generates the error to be fed to the proportional–integral controller. Then, the PI control produces the duty cycle signal which is the input to the boost converter. As a consequence, the plant model changes the operating point to follow the power command. The stability of the control system can be evaluated through the location of the poles of the closed-loop transfer function, $G_{cl}$.

For all the identified transfer functions, the poles of $G_{cl}(s)$ lay on the left half complex plane, which is a guarantee of stability. In particular, a sample of six transfer functions with diverse pole locations has been selected to further analyze the stability of the control system. Table 4 details the open-loop transfer functions selected for the stability analysis.

The closed-loop stability is confirmed with the Bode plots of the open-loop transfer function $C(s)G(s)$ in Fig. 6, where the minimum phase margin is 110 degrees and the gain margin is infinite in every case.

As explained in the previous section, in case of major irradiance drops, MPPT operation must be activated. As the tracking of the MPP is traditionally made by voltage control, the control system must be able to alternate between the control of PV power and PV voltage. This can be achieved by a switched parallel PI control diagram, depicted in Fig. 7.

As illustrated in Fig. 7, the control system is formed by two parallel PI controllers: one for the PV power control loop and the other one for the PV voltage control loop. Table 5 details the parameters of the controllers of the system under study. Both proportional–integral controllers include an external reset signal, so the error is initialized when the mode of operation is changed. It is worth noting that a smooth operation is guaranteed as the output of each loop is selected as the initial condition between modes of operation. The operation mode is defined by the signal *Mode*, which is commanded by the proposed PRRC algorithm.

## 3. Proposed photovoltaic power ramp-rate control

Previous methods for photovoltaic PRRC without energy storage tackle the problem in the same way: first, a measurement of the power ramp-rate is obtained and then, if the measured ramp is greater than a threshold, the MPPT algorithm is modified in order to reduce the output power of the system at the earliest. The main weakness of these procedures is that they operate on the principle of cause and effect, so when the PRRC is activated it may be too late. The proposed PRRC strategy employs a different approach: it regulates the PV power at every moment and the power reference is changed according to the maximum power ramp-rate permitted, never exceeding this ramp.

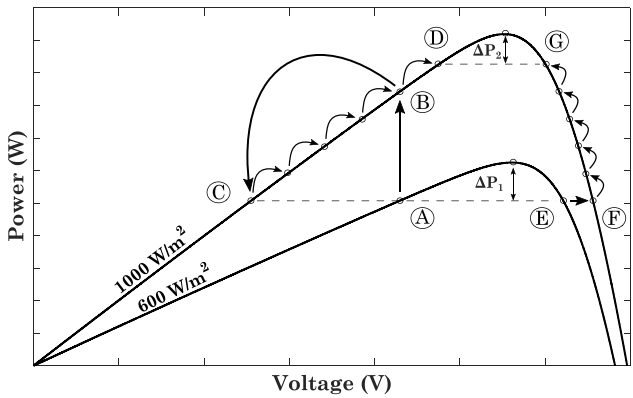

**Fig. 8.** Comparison of a previous method and the proposed one.

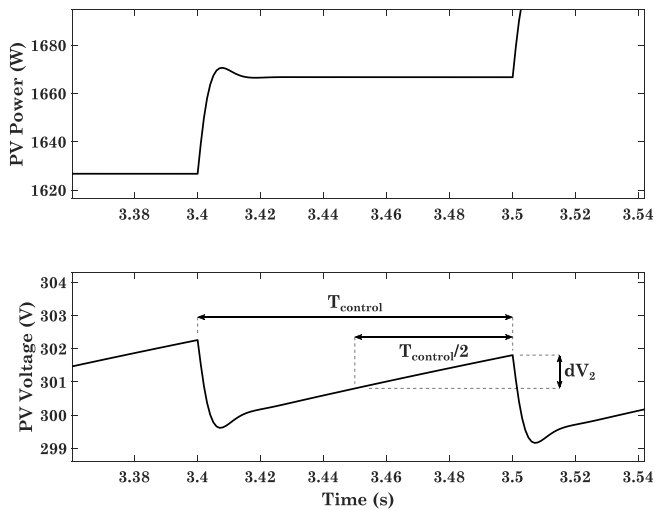

**Fig. 9.** Measurement of PV voltage variation.

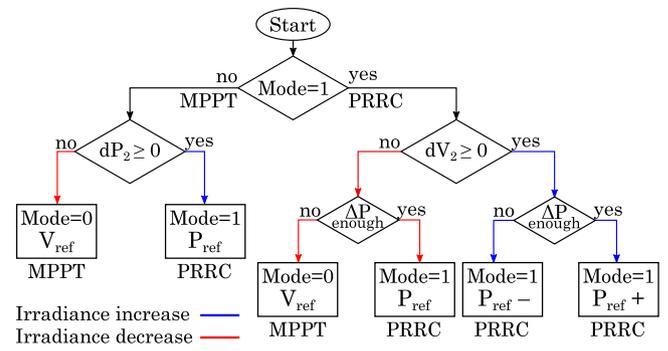

**Fig. 10.** Flowchart of the proposed PRRC.

Fig. 8 compares the method presented in [15] (left trajectory) with the proposed one (right). As depicted, both strategies start from a curtailed power level $\Delta P_1$ (points A and E, respectively). However, if irradiance rises from 600 W/m² to 1000 W/m², different responses can be observed. The voltage-regulated strategy will experience an abrupt change in power (from point A to point B) while the power-controlled method will react inherently to changes in the incident irradiance (from point E to point F). Finally, both methods modify the operating point until they reach the new curtailed level $\Delta P_2$ (points D and G, respectively). Apart from the initial abrupt change, some considerations must be taken into account: with the voltage-controlled approach, the tracking of the power reference is not trivial as it depends on the power-voltage characteristic. Furthermore, this simple analysis has been carried out with just two levels of irradiance for the sake of clarity, but, in reality, the irradiance level is continuously varying.

As explained, the proposed methodology operates the PV system at a curtailed power level with steady-state irradiance conditions. In this way, the system has an active power reserve available to smooth the output power fluctuations caused by irradiance variability. While controlling PV power, the evolution of PV voltage gives information about what is going on with the incident irradiance. In fact, at the right-hand side of the P–V curve, if PV power is kept constant, an increase in PV voltage implies an increment in irradiance. On the contrary, a PV voltage decrease signifies a reduction in the primary energy source. In order to respect the settling time of the control system, the measurement of the PV voltage variation has to be made at the second semi-period of the control period, as illustrated in Fig. 9. A similar procedure is used when MPPT mode is activated for the determination of the PV power variation, $dP_2$.

Fig. 10 shows the flowchart of the proposed PRRC strategy. The algorithm first verifies the mode of operation: if MPPT mode is activated, $dP_2$ reflects changes in irradiance. Two possibilities are considered: if irradiance is decreasing, the MPPT mode continues being enabled and the control reference is the voltage $V_{ref}$; if irradiance is increasing or maintained, PRRC is activated in search of the commanded power reserve. On the other hand, if PRRC mode is on, $dV_2$ indicates irradiance variations. Then, depending on the available power reserve $\Delta P$, four cases are contemplated: irradiance is decreasing and there is no available power reserve, so MPPT mode is activated; irradiance is decreasing but there is still power reserve, in this case the PRRC mode continues being enabled for maintaining the reference power; irradiance is increasing and there is no power reserve, so the algorithm reduces the power reference in order to follow the commanded power reserve; and irradiance is increasing and there is still power reserve, in this case, the power reference is increased according to the maximum power ramp-rate permitted.

### 3.1. Stability analysis

As the flowchart of Fig. 10 shows, switched operation modes are required for the proposed PRRC strategy. Thus, it is crucial to guarantee control stability in different operating conditions, specifically in ramp-down irradiance events while controlling PV power [26].

In this analysis, the most critical moments are considered: the transition from MPPT to PRRC mode and vice versa. When MPPT mode is activated, the operating point oscillates around the MPP without any inconvenience — this is commonly known as the three-level operation. However, if an irradiance increase is detected, the control system could establish the operating point on the left or the right-hand side of the P–V curve. The displacement of the operating point to the left of the MPP is avoided by the control system as explained in the following: if an increase in irradiance occurs, the present value of PV power ($P_{pv}$) is greater than the present value of the reference power ($P_{ref}$) and, therefore, the error that is fed to the PI power control is negative. The fact that both PI controllers have external reset signals makes the error to be re-initialized when the operation mode is switched. In this way, proportional and integral terms contribute to reducing the duty cycle, thus establishing the operating point at the right-hand side of the P–V curve.

The remaining case to be analyzed is the transition from PRRC mode to MPPT mode. The starting operating point in this case is located at the right-hand side of the P–V curve as explained above. Then, global stability and avoidance of operation at the left-hand side of the MPP must be guaranteed. In order to avoid these issues, estimation of the MPP in real time is necessary for monitoring the power reserves. Therefore, when irradiance starts to decrease, the algorithm detects a reduction of the power reserve and, before the power reserves are over, MPPT mode is enabled. Thus, stability is ensured in the most critical moments.

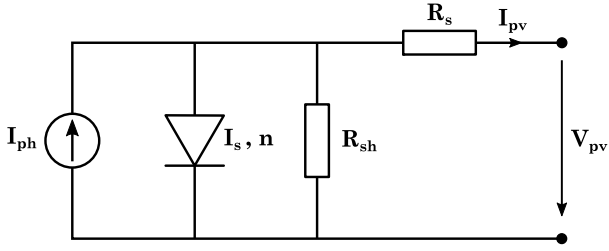

**Fig. 11.** Single-diode PV model.

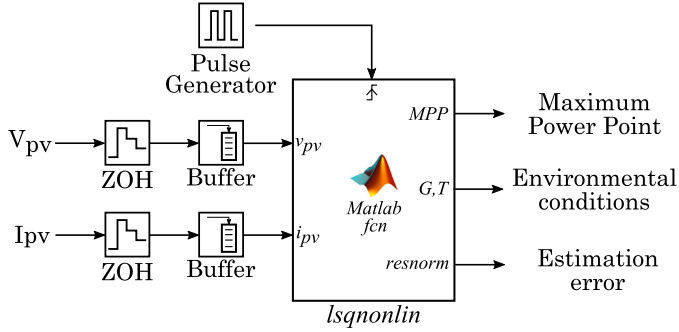

**Fig. 12.** MPP estimator implementation.

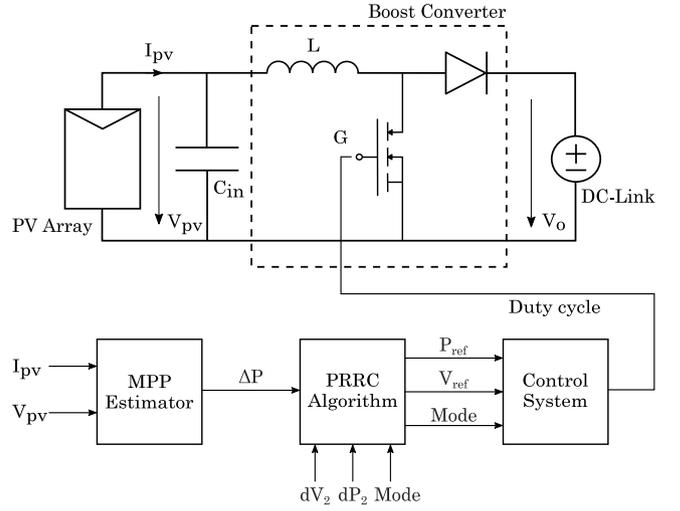

**Fig. 13.** Complete PV system with relevant signals.

### 3.2. Maximum power point estimation

The proposed photovoltaic PRRC requires information about the MPP in real-time in order to know the available power reserve. This can be achieved by the application of nonlinear least squares curve-fitting algorithms, which adjust a set of measurements to a nonlinear mathematical model. In this particular case, the mathematical model is defined by the Single-Diode Model (SDM) of the PV array, which offers a good compromise between accuracy and simplicity for allowing real-time implementation. Fig. 11 represents the equivalent circuit of the SDM together with its five electrical parameters: the photocurrent $I_{ph}$, the diode saturation current $I_s$, the diode ideality factor $n$, the shunt resistance $R_{sh}$ and the series resistance $R_s$.

It is worth mentioning that the estimation of the MPP, as well as the PV system identification (Section 2.1.1), is affected by aspects such as degradation of PV modules or limitations of the SDM. In order to minimize the impact of these drawbacks, it is recommended to recalibrate the PV system periodically, for example once per semester, considering the degradation rate indicated in [27]. An efficient way to obtain the parameters of the SDM from experimental I–V curves is described in [28]. This study shows that the discrepancies between the SDM and real field measurements are small for a wide repository of photovoltaic modules [29].

The application of the Kirchhoff's laws gives the implicit mathematical relation between the array terminal voltage and current in Eq. (3):

$$I_{pv} = I_{ph} - I_s \left[ exp\left( \frac{V_{pv} + I_{pv}R_s}{\frac{n\,kT}{q}} \right) - 1 \right] - \frac{V_{pv} + I_{pv}R_s}{R_{sh}}, \tag{3}$$

where $k$ is the Boltzmann constant, $T$ is the PV array temperature and $q$ is the electron charge. Eq. (3) can be expressed in an explicit manner through the Lambert W function [30]:

$$I_{pv} = \frac{R_s R_{sh}(I_{ph} + I_s) - V_{pv}}{R_s + R_{sh}} - \frac{V_T}{R_s} W \left[ \frac{\frac{R_s R_{sh}}{R_s + R_{sh}} I_s exp\left( \frac{R_s R_{sh}(I_{ph} + I_s) + R_{sh}V_{pv}}{V_T(R_s + R_{sh})} \right)}{V_T} \right] \tag{4}$$

The five parameters of the SDM are not constant, as they depend on the operating atmospheric conditions, i.e., the incident irradiance $G$ and the temperature $T$. De Soto et al. [31] developed the translation equation of these parameters from the Standard Test Conditions (STC: $G_0$, $T_0$) to other operating conditions:

$$I_{ph} = \frac{G}{G_0} \left[ I_{ph0} + \alpha_{1sc}(T - T_0) \right], \tag{5}$$

$$I_s = I_{s0} \left( \frac{T}{T_0} \right)^3 \exp\left( \frac{E g_0}{\frac{kT_0}{q}} - \frac{Eg}{\frac{kT}{q}} \right) \tag{6}$$

$$n = n_0 \tag{7}$$

$$R_s = R_{s0} \tag{8}$$

$$R_{sh} = R_{sh0} \frac{G_0}{G}. \tag{9}$$

where $\alpha_{1sc}$ is the short-circuit temperature coefficient, $E_g$ is the energy bandgap of the p–n junction's material and the "0" subscript indicates the parameter value at STC. Substituting Eqs. (5)–(9) into Eq. (4) gives an estimation of PV current $\hat{I}_{pv}$ based on PV voltage measurements, the reference five parameters model and the environmental conditions $(G, T)$. The objective of the curve-fitting problem is to find an estimation of a pair of values $G$, $T$ that minimize the error between a set of $m$ voltage and current measurements $(V_{pv}, I_{pv})$ and the estimated $\hat{I}_{pv}$:

$$min\, z = \sqrt{\frac{1}{m} \sum_{i=1}^{m} \left[ I_{pv,i} - \hat{I}_{pv,i}(V_{pv,i}, G, T) \right]^2} \tag{10}$$

where $z$ is the RMSE between the measurements and the estimated model. Once the environmental conditions are estimated, a single swept of the P–V curve is enough to find the estimated MPP. Finally, the available power reserve can be obtained as the difference between the estimated MPP and the actual PV power. According to [32], the estimation procedure is more accurate when the PV system is operated at the right-hand side of the MPP, which is consistent with the conclusion reached in Section 2.2.1 of this work.

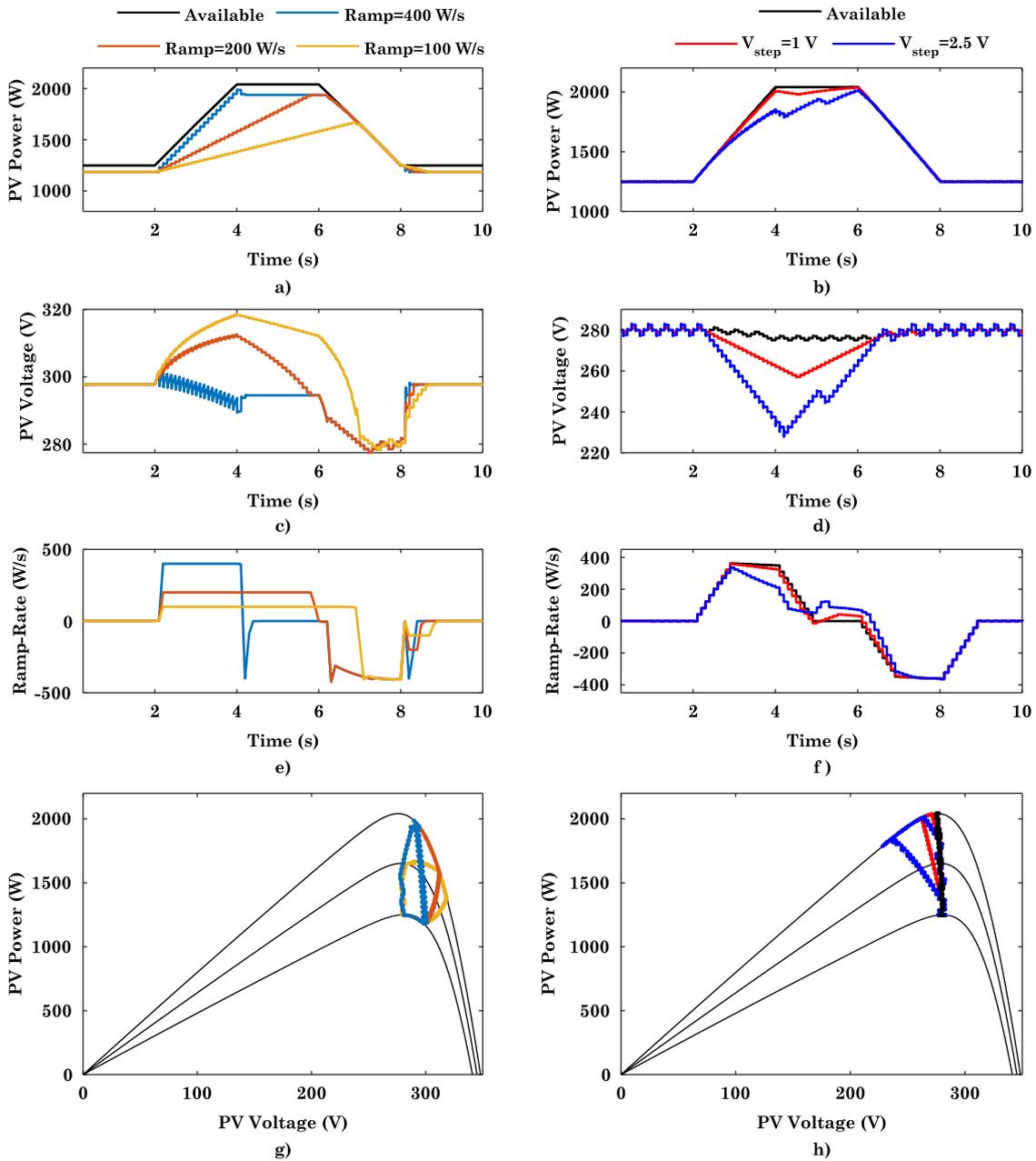

**Fig. 14.** Simulation results of the Case Study 1 for the proposed method (on the left) and the one proposed by Sangwongwanich et al. [14] (on the right): (a)–(b) PV power, (c)–(d) PV voltage, (e)–(f) Power ramp-rate and (g)–(h) Operating point trajectories over the power-voltage curve.

In this paper, the curve-fitting problem has been solved by means of the Levenberg–Marquardt method [33,34], implemented in the embedded MATLAB function *lsqnonlin*, available in the Optimization Toolbox [35]. Fig. 12 represents the implementation of the MPP estimator. As depicted, measurements of PV voltage and current are first discretized in Zero-Order Hold (ZOH) blocks with sample times of 1 ms and stored in buffers with capacity for 100 sampled data. Then, the *lsqnonlin* function is executed every 100 ms to provide real-time estimations of both MPP and environmental conditions, and the error is committed in the estimation procedure.

## 4. Simulation results

The proposed PRRC strategy has been tested in MATLAB/Simulink. Fig. 13 illustrates the complete PV system, with the main blocks and signals involved. The MPP estimator block receives a set of *m* measurements ($V_{pv}$, $I_{pv}$) and generates the power reserve signal $\Delta P$, which is fed to the PRRC algorithm block, together with the signals $dV_2$, $dP_2$ and *Mode*. As a result, the reference values $P_{ref}$, $V_{ref}$ and the mode of operation are generated and transmitted to the control system block, which finally gives the duty cycle signal to the boost converter.

Two case studies have been considered in order to verify the adequacy of the presented PRRC algorithm.

### 4.1. Case study 1

The first case study compares the response of the proposed PRRC algorithm with the one presented in [14], in the presence of a synthetic trapezoidal irradiance profile, with extreme values of 1000 and 600 W/m². As the irradiance is increased by 400 W/m² in just 2 s, three specific power ramp-rate limits have been considered for the proposed method, namely: 400, 200 and 100 W/s, with a constant power reserve of 5% of the rated capacity. On the other hand, a single power ramp-rate limit of 100 W/s has been applied for Sangwongwanich's approach, with two different values of the perturbation size ($V_{step} = 1$ V and $V_{step} = 2.5$ V) and a single value of the filtering parameter ($n = 10$). Fig. 14 depicts the performance of both methods: the proposed strategy on the left side and Sangwongwanich's method on the right. Figs. 14(a) and (b) represent the evolution of PV power in both cases, together with the maximum available power. As it is shown, the proposed algorithm can maintain active power reserves in steady-state conditions (i.e. from $t = 0$ to $t = 2$ s and from $t = 8$ to $t = 10$ s for the three ramp limits. However, the method in [14] tracks the MPP at these instants, so no active power reserve is available for compensating irradiance drops. Evolution of PV voltage is shown in Figs. 14(c) and (d). As depicted, the proposed strategy increases the operating voltage in order to control the output power for the three limits considered, while the method presented in [14] reduces the PV voltage when the power ramp-rate exceeds the limit. The power ramp-rate produced by both methodologies is represented in Figs. 14(e) and (f). It is noteworthy that the power ramp-rate is computed as given by Eq. (11), in which the filtering parameter for the proposed method is $n = 1$, so the delay in the ramp-rate calculation is minimized:

$$r(t) = \frac{P_{pv}(t) - P_{pv}(t - T_{control})}{n\, T_{control}} \qquad (11)$$

**Table 6**
Ramp-rate control metrics of the Case Study 1.

| Method | Ramp-rate limit (W/s) | Max. Ramp (+) (W/s) | Max. Ramp (−) (W/s) | Number of violations | Average of curtailment (%) |
|---|---|---|---|---|---|
| Proposed | 400 | 400.0 | −400.0 | 0 | 4.0 |
| Proposed | 200 | 200.0 | −400.0 | 1 | 8.5 |
| Proposed | 100 | 100.0 | −400.0 | 1 | 14.7 |
| Sangwongwanich (2.5 V) | 100 | 335.4 | −363.3 | 3 | 2.6 |
| Sangwongwanich (1 V) | 100 | 358.6 | −362.5 | 2 | 0.6 |

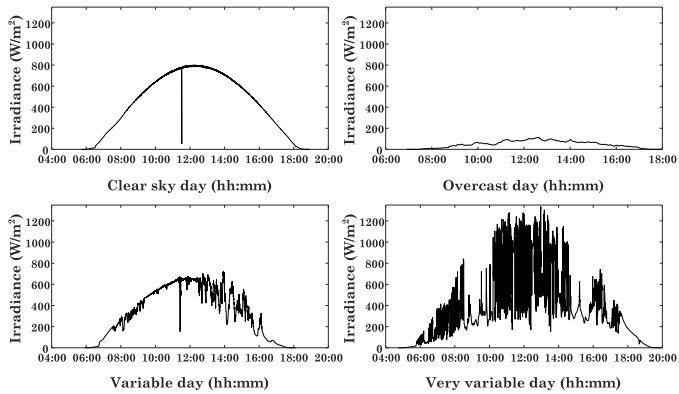

**Fig. 15.** Categories of days identified by NRCAN.

It can be seen how the proposed strategy effectively controls the power ramp-rate when the incident irradiance is increasing whereas the Sangwongwanich's method is not capable of controlling the rising power ramp-rate, neither with $V_{step} = 1$ V nor with $V_{step} = 2.5$ V. This is due to the fact that the irradiance profile used here is much more demanding than the one presented in [14], where the time scale is ten times longer and the irradiance change is ten times smaller. When irradiance drops, the proposed algorithm maintains the PV power if there is power reserve available (orange and blue curves, from $t = 6$ to $t = 6.2$ s) until this reserve is over and the MPPT mode is activated. Figs. 14(g) and (h) represent the operating point's trajectories over the power-voltage plane. As depicted, the Sangwongwanich's approach cannot reduce the PV voltage fast enough in order to meet the power ramp-rate requirements. An alternative could be to further increase the value of $V_{step}$ or decrease $n$. However, while the former would increase the oscillations in steady state, the latter would provide a more inaccurate calculation of the power ramp-rate.

Table 6 summarizes the ramp-rate control metrics of the Case Study 1. From the results, it can be inferred that the proposed method is capable of effectively control the ramp-up events and partially limit the occurrences of ramp-rate violations, compared to the Sangwongwanich's approach. Of course, the proposed strategy has an associated cost in form of average power curtailment. With respect to the negative power ramp-rates, in this case, the proposed method is not able to limit them, as the power reserves are not enough to deal with this high demanding irradiance drop. Case Study 2 presents an analysis on how these reserves can affect the power ramp-down limitation with the proposed algorithm.

### 4.2. Case study 2

The second case study considers real-field irradiance data with high temporal resolution [36]. The datasets collected by the National Resources of Canada (NRCAN) provide valuable information for short-term studies, with a sampling frequency up to 10 ms. They identify four categories of days based on cloud cover, namely: clear sky, overcast, variable and very variable. Fig. 15 illustrates examples of these categories of days.

The present case study has been carried out with a 120-s fragment of the very variable day, in order to test the effectiveness of the proposed PRRC strategy in the worst-case scenario, considering three different levels of power reserve, namely: 5, 10 and 20% of the rated capacity and a single value of the power ramp-rate limit: 100 W/s. Fig. 16 represents the simulation results of this case study. Fig. 16(a) compares the power generated by the proposed PRRC strategy with the one of the dP-P&O [37], which is the perturb and observe algorithm that best performs in rapidly changing irradiance conditions [38]. As depicted, the PRRC strategy is able to maintain different active power reserve levels in contrast with the MPPT algorithm. The power ramp-rate experienced by every case is represented in Fig. 16(b). It can be seen how the proposed method controls effectively the power ramp-up events along the simulation independently of the power reserve level. For the ramp-down events, it is also seen that as the power reserve increases, the number of ramp-rate violations is reduced. In fact, with 20% of reserves, the number of violations is zero.

Fig. 16(c) and (d) illustrate the evolution of two of the variables used for the PRRC algorithm, namely: PV voltage variation and the available power reserve. The mode of operation is represented in Fig. 16(e), being 1 the PRRC mode and 0 the MPPT one. It can be seen that the number of times that the PV system enters in MPPT mode is reduced as the power reserves increase. In fact, for the case with 20% of reserves, PRRC mode is enabled for the entire period. Finally, Fig. 16(f) depicts the real irradiance profile and the estimated one for the case of 5% of power reserves, together with the error of the estimation procedure.

**Table 7**
Ramp-rate control metrics of the Case Study 2.

| Method | Ramp-rate limit (W/s) | Max. Ramp (+) (W/s) | Max. Ramp (−) (W/s) | Number of violations | Average of curtailment (%) |
|---|---|---|---|---|---|
| dP-P&O | N/A | 827.0 | −321.5 | 28 | 0.0 |
| Proposed (5% res.) | 100 | 100.0 | −321.5 | 11 | 10.0 |
| Proposed (10% res.) | 100 | 100.0 | −264.4 | 3 | 14.2 |
| Proposed (20% res.) | 100 | 100.0 | −100.0 | 0 | 23.2 |

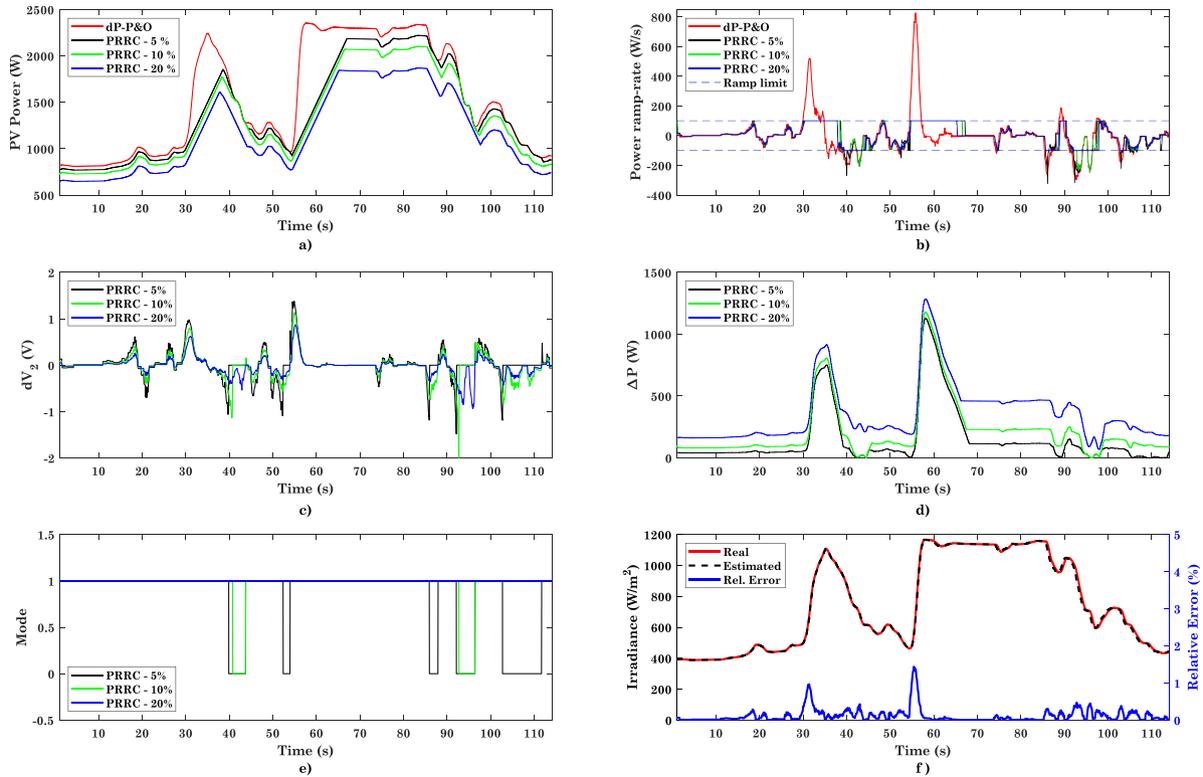

**Fig. 16.** Simulation results of Case Study 2: (a) PV power of PRRC vs dP-P&O, (b) Power ramp-rate of PRRC vs dP-P&O, (c) Variation of $V_2$, (d) PV power reserve, (e) Operation mode and (f) Irradiance: data vs estimated.

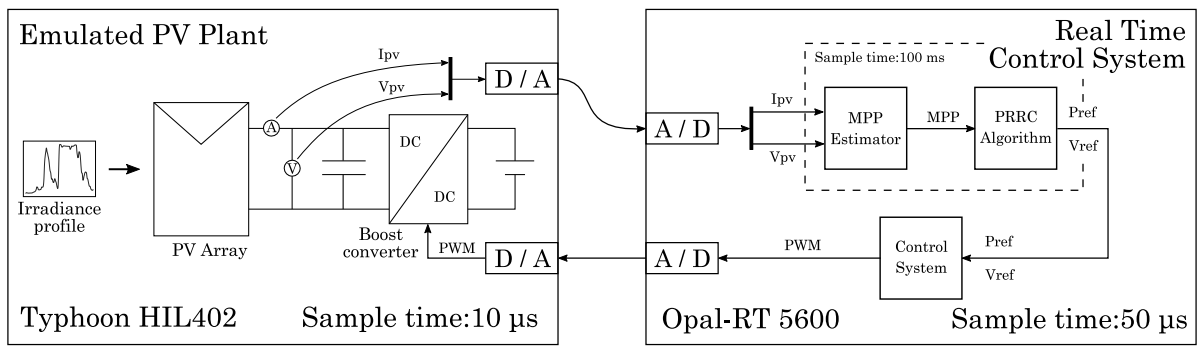

**Fig. 17.** C-HIL Management: relevant signals and sample times.

Table 7 summarizes ramp-rate control metrics of Case Study 2. The results manifest that the proposed method can limit the power ramp-rate fluctuations experienced by traditional MPPT algorithms in case of real-field high variable irradiance conditions. This Case Study also shows that, with different power reserve levels, the PRRC performance can be improved in terms of the maximum negative power ramp-rate experienced and the number of power ramp-rate violations. Finally, the results show the average curtailment applied in each case.

## 5. Experimental validation

The proposed methodology has been validated experimentally via Controller Hardware-in-the-loop (C-HIL) technique. C-HIL combines numerical simulations with classic laboratory tests. In particular, the plant to be controlled and the control system are separated into different devices, which exchange information using just analog signals in an asynchronous mode. Therefore, both devices must be Real Time Control System (RTCS) [39]. In this work, the components of the PV plant are emulated in Typhoon HIL and the control system is implemented in Opal-RT. Thus, both parts are executed in independent real-time systems. This environment needs the models and the controllers to be properly discretized. In addition, it should be noted that C-HIL implementation uses real noisy measurements and there are real possible delays in the sending of signals, resulting in procedures very close to reality. Fig. 17 schematically depicts the testing setup with relevant sample times and signals.

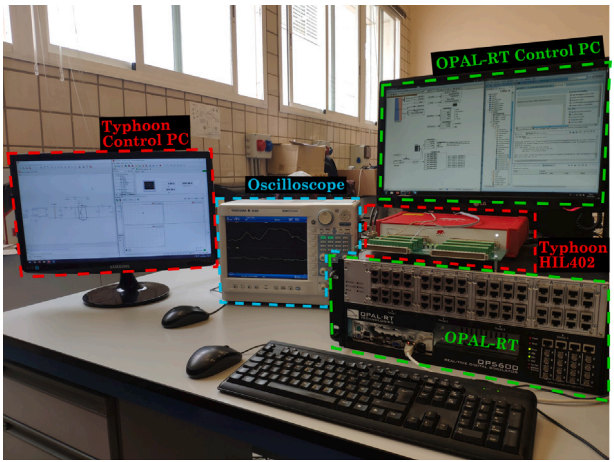

**Fig. 18.** Experimental setup.

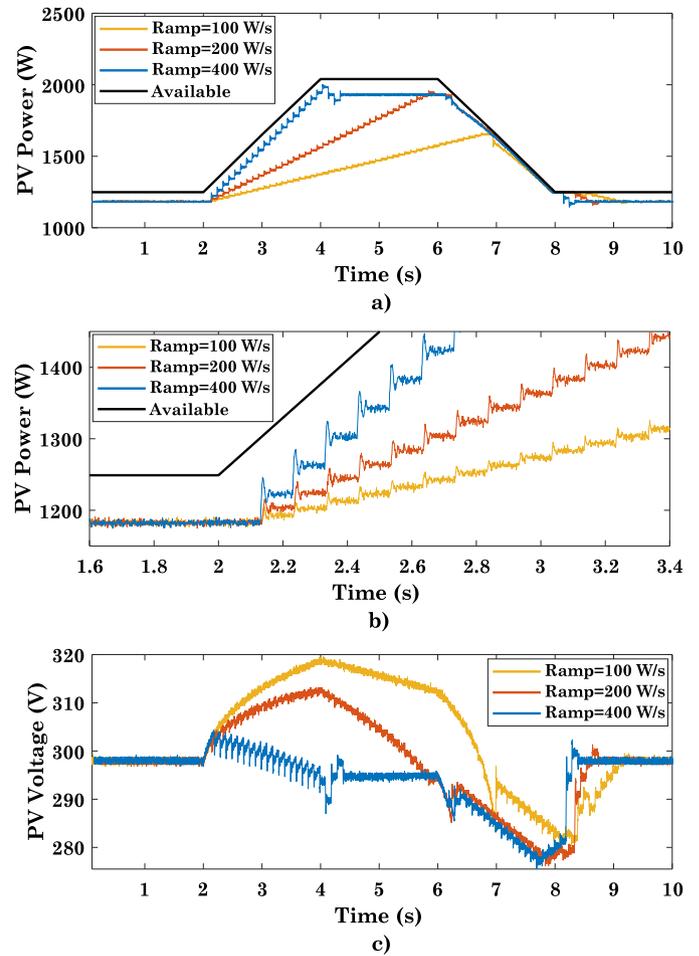

**Fig. 19.** Experimental results of Case Study 1: (a) PV power, (b) Detailed PV power and (c) PV voltage.

As shown in Fig. 17, Typhoon HIL402 runs every 10 microseconds and exchanges current and voltage measurements of the PV system through analog signals (±10 V). On the other hand, the control system runs in Opal-RT with a sample time of 50 microseconds, although the execution frequency of the MPP estimator and the PRRC algorithm is 100 ms. This system provides the PWM signal to the boost converter. Consequently, the control system just receives analog measurements of PV voltage ($V_{pv}$) and current ($I_{pv}$) and it just provides the PWM signal to the plant. Fig. 18 shows a picture of the laboratory setup, consisting of Typhoon HIL402 hardware, Opal-RT 5600 hardware, two computers for controlling Thyphoon and Opal softwares, respectively, and an oscilloscope in order to visualize and record the analog waveforms of the experiments.

### 5.1. Case study 1

The Case Study 1 has been replicated in the laboratory, with the same trapezoidal irradiance profile and considering the same power ramp-rate limits, i.e. 100, 200 and 400 W/s. Fig. 19(a) depicts the analog signals of PV power for the three ramp-rate limits applied. As shown, the experimental results are similar to those obtained by simulation. Again, the proposed PRRC algorithm is able to limit effectively the power ramp-rate in each case and to maintain active power reserves from $t = 0$ to $t = 2$ and from $t = 8$ to $t = 10$ s.

The detail of PV power evolution can be observed in Fig. 19(b). It can be seen that, at $t = 2.1$ s, an increase in irradiance is detected and the power reference is modified according to the maximum ramp-rate permitted in each case. The most noticeable difference with the simulations of Section 4 is the presence of noise in measurements.

Fig. 19(c) represents the measured PV voltage. As depicted, in every case the evolution of PV voltage is different in order to meet the requirements of power ramp-rate limits.

### 5.2. Case study 2

The Case Study 2 of section IV has been replicated in the laboratory with the same irradiance profile, considering the same power ramp-rate limit, i.e. 100 W/s and a power reserve level of 5%. Fig. 20 records the results of the experiment in an empirical version of Fig. 15. The measured analog signal PV power is depicted in Fig. 20(a) together with the estimation of MPP. Fig. 20(b) represents the power ramp-rate of the mentioned waveforms. As depicted, whereas the proposed method can effectively adjust the power ramp-up to the limit, the ramp-rate of the estimated MPP exceeds the limit twice (at $t = 17$ and $t = 42$ s).

The evolution of the analog signal $dV_2$ is illustrated in Fig. 20(c) with its associated digital signal used in the PRRC algorithm. It is worth noting that the analog signal is filtered with a 500 Hz filter. Fig. 20(d) depicts the available power reserve, computed as the difference between the MPP and the PV power. The operation mode is shown in Fig. 20(e), where $Mode = 1$ refers to PRRC mode and $Mode = 0$ refers to MPPT. Fig. 20(f) represents the real irradiance profile, the estimated one and the estimation error committed.

## 6. Conclusion

In this paper, a novel storageless photovoltaic Power Ramp-Rate Control is presented. Compared to the existing methods in the literature, the proposed algorithm regulates PV power rather than PV voltage, which makes the PV system react inherently to sudden changes in the incident irradiance. Another advantage of the methodology presented is that it is able to adjust power reserves as required in every moment. In this way, when the objective is to minimize the energy waste, a low power reserve level or even zero (MPPT) can be applied. On the other hand, if the objective imposes to never violate the power ramp-down limit, the power reserve can be set at a higher percentage of the rated capacity. Logically, the latter situation should not be prolonged for a very long period, as the losses would be unacceptable, but it is possible in short periods of time in which irradiance is highly variable.

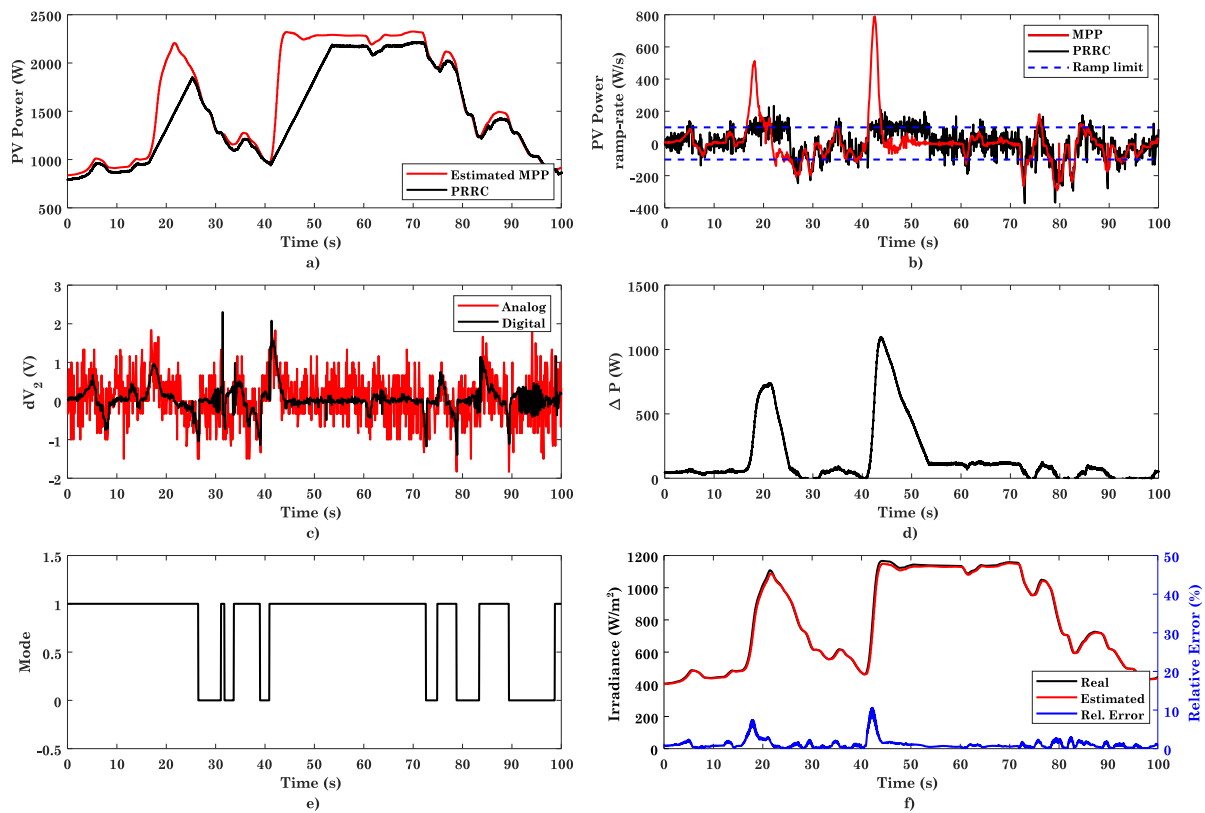

**Fig. 20.** Experimental results of Case Study 2: (a) PV power vs estimated MPP, (b) Power ramp-rate of PRRC vs MPP, (c) Analog variation of $V_2$ vs digital signal, (d) PV power reserve, (e) Operation mode and (f) Irradiance: data vs estimated.

The plant model used for the design of the control system has been obtained by direct system identification from sampled data, therefore avoiding linearization processes. The presented method has been theoretically justified and tested in simulation, firstly with a synthetic irradiance profile considering different power ramp-rate levels and finally with real-field highly variable irradiance data and several power reserve levels. The method has been compared to another PRRC strategy based on power curtailment, demonstrating the effectiveness of the presented PRRC algorithm. The results also show that a proper power ramp-down limitation depends on the previous power reserve level applied. Although this implies an energy waste, it can be used at specific moments in which irradiance is highly variable. Experimental validation through a Controller Hardware-in-the-loop methodology has been carried out. As a result, the feasibility of the proposed algorithm has been validated, not being affected by noisy measurements or transmission delays.

## CRediT authorship contribution statement

**Jose Miguel Riquelme-Dominguez:** Conceptualization, Data curation, Formal analysis, Investigation, Methodology, Resources, Software, Validation, Visualization, Writing – original draft, Writing – review & editing. **Francisco De Paula García-López:** Resources, Software, Validation, Visualization, Writing – review & editing. **Sergio Martinez:** Conceptualization, Formal analysis, Funding acquisition, Investigation, Methodology, Project administration, Resources, Supervision, Validation, Writing – review & editing.

## Declaration of competing interest

The authors declare that they have no known competing financial interests or personal relationships that could have appeared to influence the work reported in this paper.

## Acknowledgments

This research was funded by the Spanish national research agency Agencia Estatal de Investigación, grant number PID2019-108966RB-I00 /AEI/ 10.13039/501100011033.